\documentclass[12pt]{article}
\usepackage{pic03}
\usepackage{hyperref}
\usepackage{url}
\usepackage{graphicx}

\begin{document}

\title{\bf RELEVANT RESULTS FROM THE NA48 EXPERIMENT}
\author{
E. Imbergamo        \\
{\em Department of Physics, University of Perugia, Italy} \\
M. Piccini        \\
{\em Department of Physics, University of Perugia, Italy} \\
M.C. Petrucci       \\
{\em INFN, section of Perugia, Italy}}
\maketitle

\baselineskip=14.5pt
\begin{abstract}
We report relevant results from NA48 experiment at CERN SPS. NA48 was proposed
in 1990 \cite{proposal} to 
study direct CP violation in $K^0\rightarrow\pi\pi$ to a level of accuracy
sufficient to resolve the inconclusive status left by the previous measurements
performed by NA31 \cite{NA31} and E731 \cite{E731}.
In 2002 NA48 published the final result 
\cite{NA48epsoeps}. Small modification to the experimental setup have allowed
NA48 to go forward with an extensive investigation of 
$K^0$ rare decays and hyperon decays. Some results are already available 
and reported here together with the final CP violation measurement.
\end{abstract}

\baselineskip=17pt

\section{The measurement of $Re(\epsilon^\prime/\epsilon)$}
\begin{figure}              
\begin{center}              
  \includegraphics[scale=0.40]{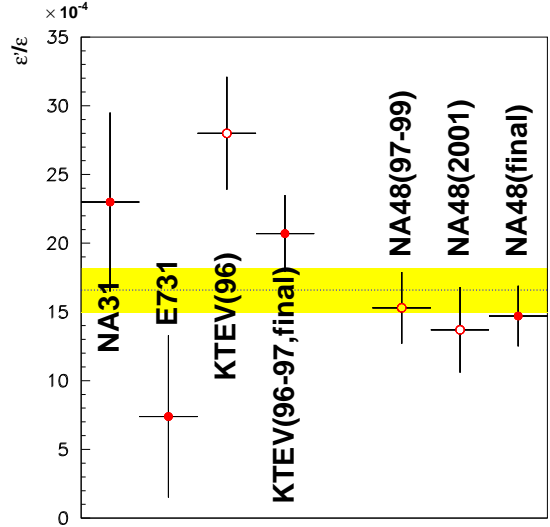} 
 \caption{\it
   World measurements of $Re(\epsilon^\prime/\epsilon$)
    \label{eoe} }
\end{center}
\end{figure}

It is well known that the decay of $K^0/\overline{K^0}$ 
into two pions violates CP. Such a violation can have two contributions: 
one, indirect, associated to $K^0/\overline{K^0}$ mixing and another,
direct, coming from the decay amplitude. The amount of direct CP violation
in this decay is parametrized by the parameter $Re(\epsilon^\prime/\epsilon)$,
 which can be computed 
in the framework of the standard electro-weak-model, albeit with large 
theoretical uncertainties.     
Typical theoretical predictions of $Re(\epsilon^\prime/\epsilon)$
varies from few $10^{-4}$ to about $2\times10^{-3}$, even though with large
exceptions. \\
$Re(\epsilon^\prime/\epsilon)$ is connected to the double ratio of decay rates
according to the following formula:
\begin{equation}
R =  
\frac{\Gamma(K_L\rightarrow\pi^0\pi^0)}{\Gamma(K_S\rightarrow\pi^0\pi^0)}/
\frac{\Gamma(K_L\rightarrow\pi^+\pi^-)}{\Gamma(K_S\rightarrow\pi^+\pi^-)} 
\simeq 1 - 6\cdot Re(\epsilon^\prime/\epsilon) \\
\end{equation}
In order to exploit the cancellation in the double ratio of systematic 
uncertainties, the experimental apparatus consists of two concurrent and
almost co-linear beams, one providing  the experiment with $K_L$ decays 
and the other one with $K_S$ decays and 
the $K^0$ to $\pi\pi$ decays are reconstructed in the same decay region.
A tagging station is devoted to identify the $K^0$ decay as a $K_S$ or a $K_L$
decay. 
In 2002 NA48 has published \cite{NA48epsoeps} the final result
of the measurement:
\begin{equation}
    \large
    Re(\epsilon^\prime/\epsilon) = (15.3\pm2.6)\times 10^{-4}
\end{equation}
It is the most precise measurement ever done as shown
in Fig.\ref{eoe}. The world average value is also reported 
together with the uncertainty (yellow band). 
\section{Selected items about rare decays}
The radiative kaon decays are ideally suited to test the validity 
of the Chiral Perturbation Theory ($\chi PhT$). 
The processes may be described in a perturbative expansion 
of momenta: ${\cal O}(p^2),{\cal O}(p^4)$.
Examples of this type are 
$K_S\rightarrow\gamma\gamma$ and $K_L\rightarrow\pi^0\gamma\gamma$.
In both cases there is no contribution from the ${\cal O}(p^2)$ term, 
while the ${\cal O}(p^4)$ contribution is predicted to better than 5$\%$ 
by $\chi PhT$. NA48 has measured \cite{ksgg},\cite{klpgg}:
\begin{equation} 
BR(K_S\rightarrow\gamma\gamma)=(2.78\pm 0.06_{stat}\pm 0.02_{MCstat}
\pm 0.04_{syst})\times 10^{-6}
\end{equation} 
\begin{equation} 
BR(K_L\rightarrow\pi^0\gamma\gamma)=(1.36\pm 0.03_{stat}\pm 0.03_{syst}
\pm 0.03_{norm})\times 10^{-6}
\end{equation} 
%
%\begin{figure}[t]              
%\begin{center}              
%  \includegraphics[scale=0.50]{plots/ksgg.eps} 
% \caption{\it
%   World measurements of $BR(K_S\rightarrow\gamma\gamma)$ and $\chi PhT$
%   prediction.
%    \label{ksgg}}
%\end{center}
%\end{figure}              
%
The value of $BR(K_S\rightarrow\gamma\gamma)$ deviates 
from ${\cal O}(p^4)$ prediction and indicates a large ${\cal O}(p^6)$ 
contribution. The ${\cal O}(p^4)$ contribution
to $K_L\rightarrow\pi^0\gamma\gamma$ turns out to be also an underestimation
of the decay rate. Anyway at ${\cal O}(p^6)$ the rate may be reproduced 
by adding a contribution from the VDM mechanism, via the coupling constant
$a_v$ that NA48 has measured to be:
\begin{equation} 
a_v=-0.46\pm 0.03_{stat}\pm 0.04_{syst}
\end{equation} 
\section{Hyperon decays in NA48}
The target used for the production of $K^0$s is also a huge source of
hyperons.
By using the small fraction of hyperons that passed the standard triggers in 
the previous years, 
the NA48 collaboration has already published results on hyperon 
physics \cite{hypNA48}:
\begin{equation}
m(\Xi^{0})=[1314.82\pm0.06(stat.)\pm0.20(syst.)]MeV/c^{2}
\end{equation}
\begin{equation}
BR(\Xi^{0}\rightarrow\Lambda \gamma) =[1.90\pm0.34(stat.)
\pm0.19(syst.)]\times 10^{-3}
\end{equation}
\begin{equation}
BR(\Xi^{0}\rightarrow\Sigma^{0} \gamma)=[3.14 
\pm0.76(stat.)\pm0.32(syst.)]\times 10^{-3}
\end{equation}
In the 2002 special triggers have been dedicated to hyperon decays and NA48
claims main achievements to:
\begin{enumerate}
\item
study form factors and flavor symmetry violations in the $\Xi^{0}$ decays;
\item
give an alternative measurement of $V_{us}$ (CKM parameter) using $\Xi^0$ (and
$\Lambda$) beta decay instead of kaon beta decays. 
\end{enumerate}

\end{document}